\documentstyle[12pt,twoside,fleqn,espcrc1,epsf]{article}

\newcommand{\be}{\begin{eqnarray}}
\newcommand{\ee}{\end{eqnarray}}

\newcommand{\AmS}{{\protect\the\textfont2
  A\kern-.1667em\lower.5ex\hbox{M}\kern-.125emS}}

\title{Diquark Condensation in High Density Baryon Matter}

\author{Thomas Sch\"afer\address{Institute for Nuclear Theory\\
        University of Washington\\
        Seattle, WA 98195}}

\begin{document}
\maketitle

\begin{abstract}
  We argue that cold quark matter is a diquark Bose condensate. 
The Cooper pairs of QCD are spin-isospin zero, color anti-symmetric 
quark pairs. For two light flavors, instanton effects lead to gaps
on the order of 50 MeV. 
\end{abstract}

\section{Introduction}

   Over the past years, important progress has been achieved  
in our understanding of the phase structure of QCD at finite 
temperature. Even though many important points remain to be 
worked out (like the nature of the QCD phase transition for 
realistic values of the quark masses), there is a nice frame
work based on universality arguments and our ability to 
perform simulations on the lattice. 

   For reasons that we do not need to reiterate here, the problem 
of cold dense matter is much less understood. On the other hand, 
it has been realized for quite some time that the possible phase 
structure of dense matter is very rich. In addition to the nuclear 
and quark matter phases, new phases containing pion or kaon condensates, 
strange quark matter, etc., have been suggested. In this contribution 
we want to study the possibility that cold quark matter is in a 
superconducting phase. The Cooper pairs of QCD are spin zero diquarks. 

 Unlike many of the phases that we just mentioned, this phenomenon is 
very robust and independent of the detailed dynamics. It is based on
the observation that a sharp Fermi surface is expected to be unstable 
with respect to pair condensation whenever there is an (arbitrarily weak!) 
attractive interaction between quarks pairs in the vicinity of the 
Fermi surface.

\section{The Quark-Quark interaction}

  While the phenomenon as such is independent of the strength 
and the exact form of the interaction, the size of the gap, the 
condensation energy, the critical temperature etc., certainly
depend on the interaction. If the chemical potential is very 
large, we expect the interaction to be perturbative. The 
Coulomb interaction between quarks is attractive if the two
quarks are in a spin-isospin zero, color anti-triplet state.
Color superconductivity induced by perturbative gluon exchange
was first studied in detail by Bailin and Love \cite{BL}
(the possibility of superconductivity in cold quark matter 
was apparently first pointed out by Frautschi \cite{Frautschi}). 
These authors find that at baryon densities $\rho\sim (5-10) 
\rho_0$, where $\rho_0$ is nuclear matter density, both the 
gap and the critical temperature are on the order of 1 MeV. 

  In this contribution we show that non-perturbative effects can 
lead to diquark condensates with $\Delta,T_c$ about two orders of 
magnitude bigger \cite{ARW_97,RSSV_97}. These non-perturbative 
effects are connected with instantons. For two flavors (up and 
down) the $(\bar qq)$ interaction generated by instantons is given 
by \cite{tHooft}
\be
\label{l_inst}
{\cal L} &=& G\frac{1}{4(N_c^2-1)}
\left\{ \frac{2N_c-1}{2N_c}\left[
     (\bar\psi\tau_\alpha^{-}\psi)^2 +
     (\bar\psi\gamma_5\tau_\alpha^{-}\psi)^2\right] 
  + \frac{1}{4N_c}(\bar\psi\sigma_{\mu\nu}\tau_\alpha^{-}\psi)^2 \right\},
\ee
where $N_c$ is the number of colors and $\tau^-=(\vec\tau,i)$ is an 
isospin matrix. Phenomenology (or interacting instanton calculations)
gives $G\simeq 490\,{\rm GeV}^{-2}$. There is a great deal of evidence
that this simple interaction correctly describes many aspects
of hadronic phenomenology, like chiral symmetry breaking and correlation
functions of hadronic currents. See \cite{SS_97} for a review of these 
issues. 


\begin{figure}[t]
\begin{minipage}[t]{80mm}
\epsfxsize=7cm
\epsffile{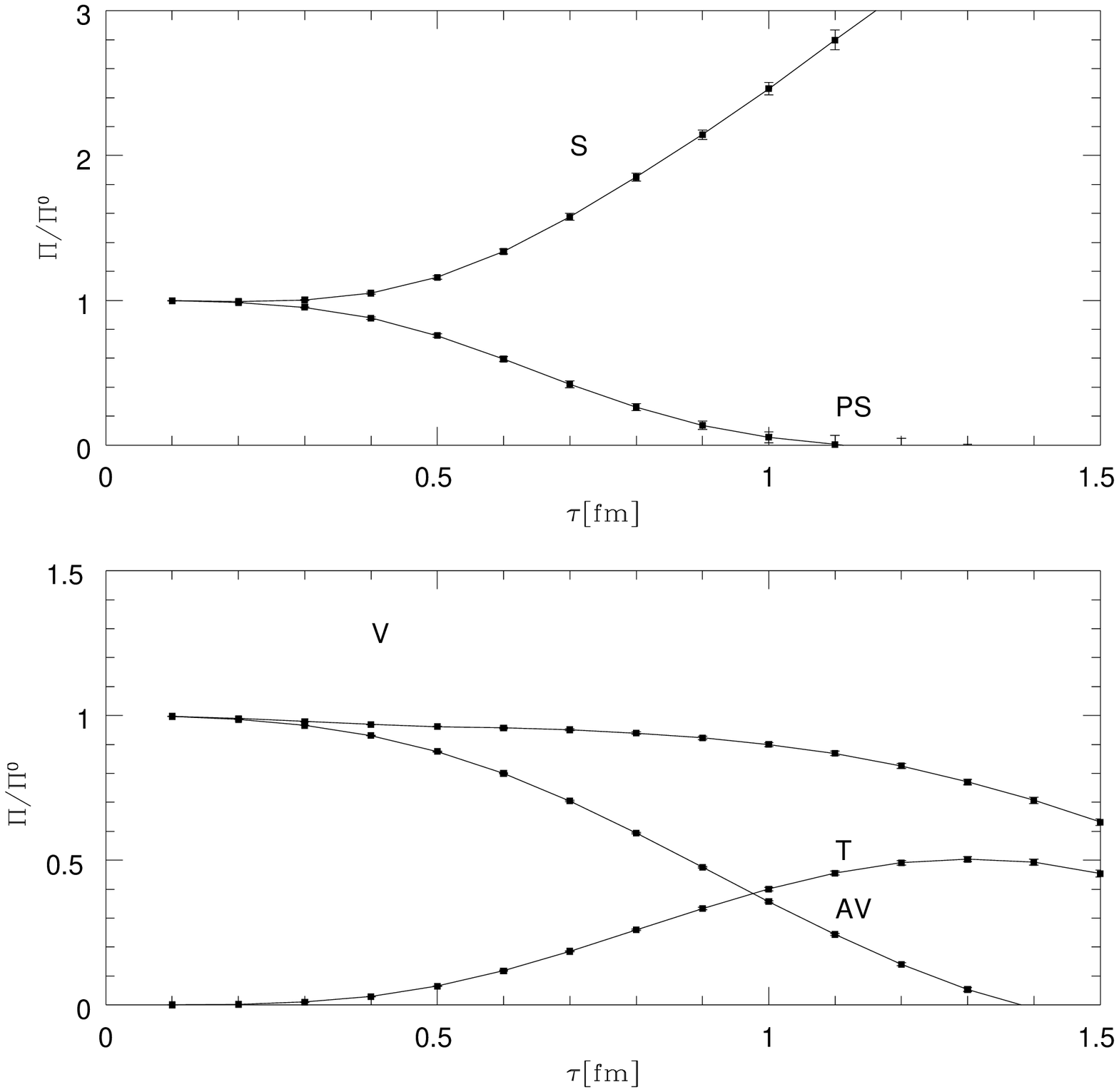}
\vspace*{-1cm}
\caption{\label{fig_diq}
Color $\bar 3$ diquark correlation functions.}
\end{minipage}
\hspace{\fill}
\begin{minipage}[t]{75mm}
\epsfxsize=7cm
\epsffile{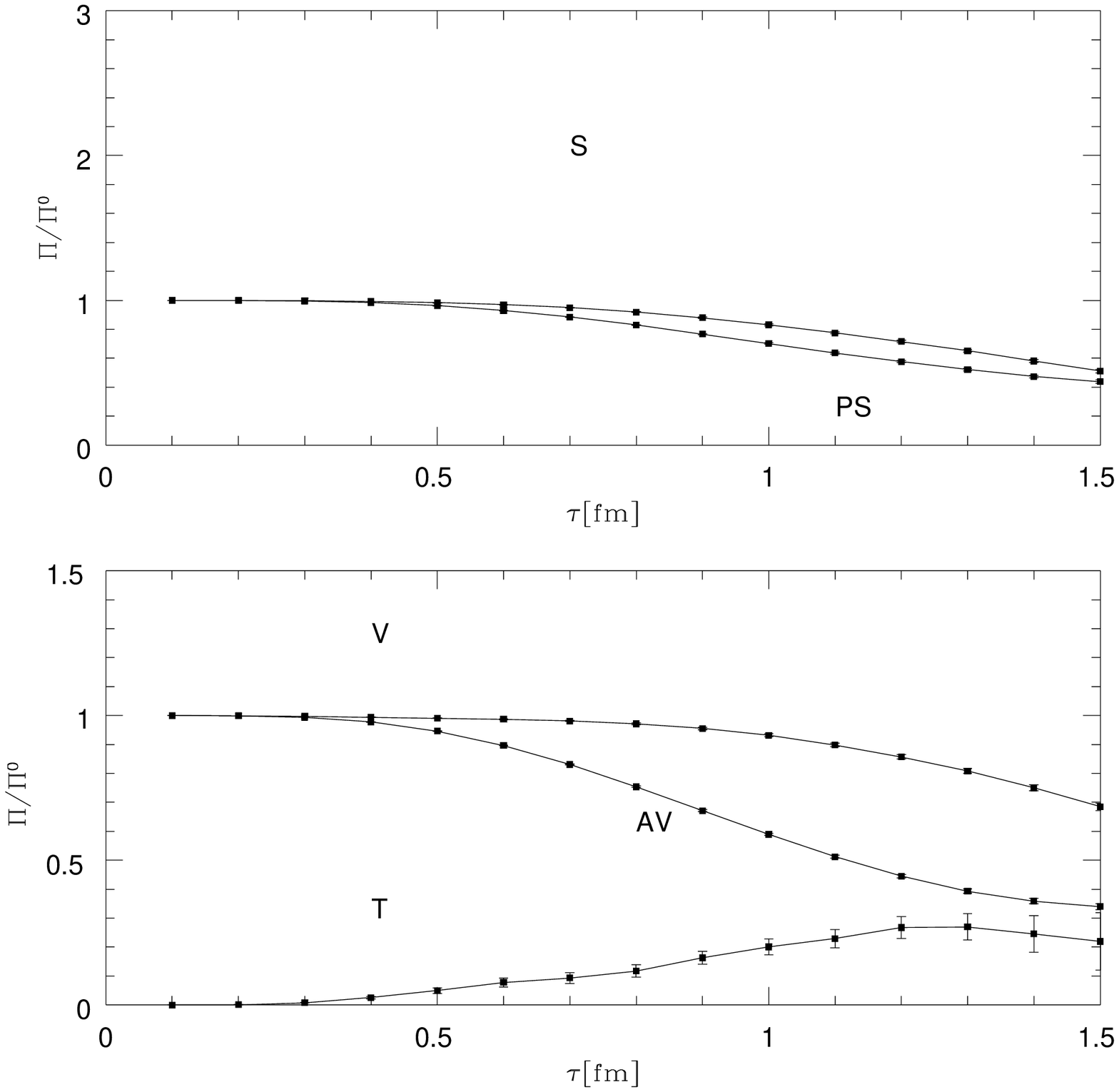}
\vspace*{-1cm}
\caption{\label{fig_diq2}
Color $6$ diquark correlation functions.}
\end{minipage}
\end{figure}

  The result (\ref{l_inst}) can be Fierz-rearranged into a $(qq)$ 
interaction. We find
\be
\label{l_diq}
{\cal L} &=&
G \left\{
 -{1\over 16 N_c (N_c-1)}
  \left[ (\psi^T C \tau_2 \lambda_A^a \psi)
         (\bar\psi\tau_2 \lambda_A^a C \bar\psi^T)
          +(\psi^T C \tau_2 \lambda_A^a \gamma_5 \psi)
           (\bar\psi \tau_2 \lambda_A^a \gamma_5 C \bar\psi^T) \right]
              \right. \nonumber\\
    & &  \left. \hspace{-0.5cm}\mbox{}
      \!   \! +{1\over 32 N_c (N_c+1)}
           (\psi^T C \tau_2 \lambda_S^a \sigma_{\mu \nu} \psi)
           (\bar\psi \tau_2 \lambda_S^a \sigma_{\mu \nu} C \bar\psi^T)
           \right\} \nonumber,
\ee
Here, $C$ is the charge conjugation matrix, $\tau_2$ is the anti-symmetric 
Pauli matrix, $\lambda_{A,S}$ are the anti-symmetric (color $\bar 3$) and 
symmetric (color 6) color generators. The effective lagrangian (\ref{l_diq}) 
provides a strong attractive interaction between an up and a down quark with 
anti-parallel spins ($J^{P}=0^{+}$) in the color anti-triplet channel, and
a repulsive interaction in the $0^-$ channel. 

  The scalar diquark current $j^a_S=\epsilon_{abc}q^T_bC\gamma\tau_2q_c$
is of course not gauge invariant, and does not couple to physical states.
But we can neutralize color by adding an infinitely heavy quark, and 
consider correlation functions of the gauge invariant current $j=j^a_SQ^a$.
In the limit $m_Q\to\infty$, the propagator of the heavy quark reduces
to a gauge string. 

  The instanton liquid does not confine, and we can calculate the 
mass of two-quark bound states. From a simple RPA calculation, we
find a scalar diquark mass of
$m_S\simeq 400$ MeV, which is significantly below the two-quark
threshold, $2m_q-m_{S}\simeq 200-300$ MeV. All other channels 
(vectors and axial-vectors, color 6 diquarks, etc.) are at most 
very weakly bound. This result is consistent with earlier calculations
in the NJL model, see for example \cite{Vogl_90}. A calculation
of diquark masses in the instanton model was also performed in 
\cite{DFL_96}. These authors perform a simultaneous meson-diquark
bosonization of the interaction (\ref{l_inst}). In their scheme, 
only a fraction $1/N_c$ of the interaction acts in the diquark 
channel. As a result, the scalar diquark is unbound. 

  Higher order effects can be taken into account by performing
numerical calculations of diquark correlation functions in the 
instanton liquid \cite{SSV_94}. Results for scalar (S), pseudo-scalar
(PS), vector (V), axial-vector (AV), and tensor (T) correlation
functions, in both color $\bar 3$ and $6$ channels are shown in 
Figs. 1 and 2. We observe that only the color $\bar 3$ scalar 
shows substantial attraction. Numerically, we find $m_S\simeq
400$ MeV, and a scalar-vector diquark splitting of almost 500
MeV. Recently, He{\ss} et al. calculated diquark correlation 
functions (in a fixed gauge, not with the gauge string included)
on the lattice \cite{HKLW_98}. They find $m_S\simeq 650$ MeV
and a smaller scalar-vector diquark splitting, $m_V-m_S=100$ MeV. 
On the other hand, they also have a nucleon-delta splitting 
which is too small, $m_\Delta-m_N=150$ MeV, while the instanton
model tends to overestimate this quantity. So the truth is
probably somewhere in between.  

  The fact that color $\bar 3$ scalar-isoscalar diquarks are
favored is not specific to instantons. If the chemical potential
is very large, semi-classical fields are exponentially suppressed,
and the quark-quark interaction is dominated by one-gluon exchange.
The effective four-fermion vertex corresponding to one-gluon 
exchange is
\be 
\label{l_oge}
{\cal L}_{OGE} &=& 
\frac{G_V}{2} \left\{
  \frac{4}{3}\left[
        (\psi^T C\gamma_5 \tau_2 \lambda_A^a \psi)
        (\bar\psi\gamma_5\tau_2 \lambda_A^a C \bar\psi^T) 
       -(\psi^T C \tau_2 \lambda_A^a  \psi)
        (\bar\psi \tau_2 \lambda_A^a C \bar\psi^T) \right] \right. 
               \nonumber\\
 & &  \left. \mbox{}
    + \frac{2}{3}\left[
        (\psi^T C\gamma_\mu\gamma_5 \tau_2\lambda_A^a \psi)
        (\bar\psi\gamma_\mu\gamma_5\tau_2 \lambda_A^a C \bar\psi^T) 
       +(\psi^T C \gamma_\mu\tau_2\vec\tau \lambda_A^a  \psi)
        (\bar\psi \gamma_\mu\tau_2\vec\tau \lambda_A^a C \bar\psi^T) 
       \right] \right\},
  \nonumber
\ee 
which is also attractive in the scalar diquark channel. The coupling
is given by $G_V= (4\pi\alpha_s)/\Lambda^2$, where $\Lambda$ is some 
IR cutoff. In the high density phase, it seems reasonable to set 
$\Lambda$ equal to the Debye mass. 

 In the vicinity of 
the chiral phase transition, higher order instanton effects may also
play a role. The interaction induced by instanton-anti-instanton 
molecules is \cite{SSV_95}
\be 
\label{l_mol}
{\cal L}_{IA} &=& 
 G_M \left\{
  \frac{1}{6}\left[
        (\psi^T C\gamma_5 \tau_2 \lambda_A^a \psi)
        (\bar\psi\gamma_5\tau_2 \lambda_A^a C \bar\psi^T) 
       -(\psi^T C \tau_2 \lambda_A^a  \psi)
        (\bar\psi \tau_2 \lambda_A^a C \bar\psi^T) \right] \right. 
               \nonumber\\
 & &  \left. \mbox{}
    + \frac{1}{24}\left[
        (\psi^T C\gamma_\mu\gamma_5 \tau_2\lambda_A^a \psi)
        (\bar\psi\gamma_\mu\gamma_5\tau_2 \lambda_A^a C \bar\psi^T) 
       -(\psi^T C \gamma_\mu\tau_2\vec\tau \lambda_A^a  \psi)
        (\bar\psi \gamma_\mu\tau_2\vec\tau \lambda_A^a C \bar\psi^T) 
       \right] \right\} , 
  \nonumber
\ee 
which is also most attractive in the scalar diquark channel.

\begin{figure}[t]
\vspace*{-2cm}
\epsfxsize=10cm
\epsffile{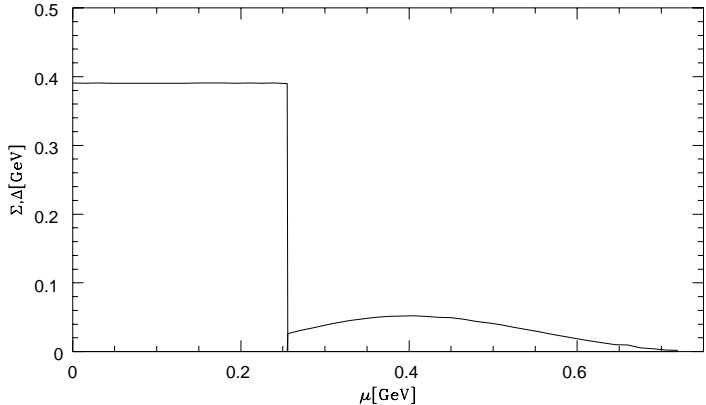}
\vspace*{-2.5cm}
\caption{\label{fig_bcs}
Chiral and diquark gaps $\Sigma$ and $\Delta$ as a function of the 
quark chemical potential.}
\end{figure}

\section{Diquark Condensation}

  We now turn to a simple mean field model of chiral symmetry 
breaking and diquark condensation. We consider the instanton 
induced effective interaction (\ref{l_inst}) and a trial 
state with both quark $\langle\bar qq\rangle$ and diquark 
$\langle q^TC\gamma_5\lambda_2\tau_2q\rangle$ condensates. 
We will denote the corresponding gaps by $\Sigma$ and 
$\Delta$. 

  At small chemical potential, quark-anti-quark condensation
is favored over diquark condensation. The size of the gap is
controlled by the standard, NJL-type gap equation. A chiral 
condensate is only formed if the interaction strength exceeds
a certain critical value. If the chemical potential is 
increased, the quark-anti-quark interaction is partially 
blocked, but the quark-quark interaction in the vicinity 
of the Fermi surface is enhanced. The diquark gap equation
in the high density phase is 
\be
\label{gap}
1=\frac{8}{(2\pi)^2} \ G(\mu) \int p^2 dp 
\frac{F((p-p_F)^2)}{(p-\mu)^2+\Delta^2 }, 
\ee
which shows the logarithmic enhancement near $p=p_F$. In the
high density phase, the coupling is exponentially suppressed,
$G(\mu)=G(0)\exp [-N_f\rho^2\mu^2\theta(\mu-\mu_c)]$. The 
quark interaction is due to instanton zero modes near the 
Fermi surface. The corresponding form factor $F$ is therefore
peaked at $p_F$. Here, we assume a simple monopole shape with 
a range $\Lambda=300$ MeV. The result for the coupled system 
of gap equations is shown in Fig. 3. At $\mu=0$ the chiral gap 
is $\Sigma=400$ MeV (this is how we fixed $G(0)$). We then find 
a first order transition to a diquark condensed state at $\mu=270$ 
MeV (see also \cite{BR_98}). The gap first grows because of the 
increase in the number of states near the Fermi surface, but
becomes very small at large chemical potential because of the
instanton suppression factor. The maximum gap $\Delta$ is on
the order of 50 MeV. 

  It is instructive to discuss the structure of the diquark 
condensed phase in terms of the corresponding Landau-Ginzburg
effective free energy
\be
\label{LG}
 F &=& \alpha \left|(\Delta^a)^2\right| 
     + \beta  \left|(\Delta^a)^2\right|^2
     + \gamma \left|(D_\mu\Delta)^2\right| 
     + \frac{1}{4}\,G_{ij}^{\, 2} 
     + \ldots . 
\ee
The coefficients $\alpha$ and $\beta$ can be matched to the 
gap equation (\ref{l_diq}), while $\gamma$ can be determined 
by performing a derivative expansion. From the 
effective action, we can read off the condensation 
energy\footnote{This sounds small, but it is not so 
different from the chiral condensation energy $\epsilon
\simeq -f_\pi^2\Sigma^2/2 \simeq -75\,{\rm MeV}/{\rm fm}^3$.}
$\epsilon=-30\,{\rm MeV}/{\rm fm}^3$ and the coherence length 
$\xi_\Delta=(\gamma/\alpha)^{1/2}\simeq 0.8$ fm. Among the eight 
gluons, 5 get a mass via the Higgs mechanism (corresponding
to breaking color $SU(3)$ to $SU(2)$). The mass of the 
off-diagonal gluons is $m_g = (\gamma/2)^{1/2}g\Delta$.
From this, we get a penetration depth $\xi_A \simeq 1.5$ 
fm. The order parameter carries electric charge, but the 
diagonal gluons mix with the photon, producing a new 
massless gauge boson. The corresponding Weinberg angle 
is small, $\tan\theta = e/(\sqrt{3}g)$. 

  It is unfortunate that, at least in the case of two flavors, 
the effective action does not support any textures. We should note
that, since the condensation energy is small, the critical 
color-magnetic field is also small, $B^2\simeq (130\, {\rm MeV})^2$.
This is significantly smaller than the field inside an instanton,
which implies that the order parameter is probably very 
inhomogeneous. 

\begin{figure}[t]
\begin{minipage}[t]{80mm}
\epsfxsize=7cm
\epsffile{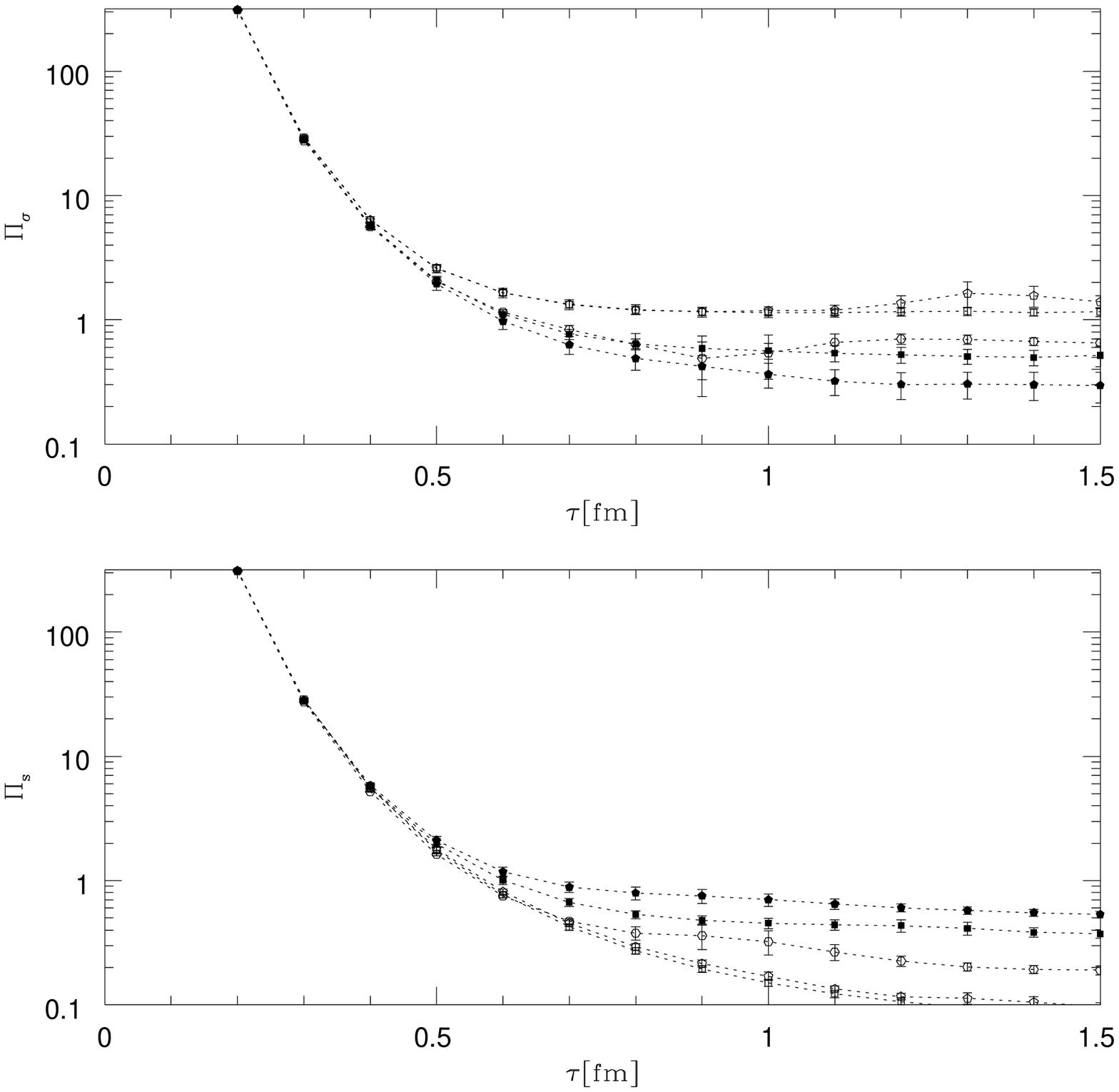}
\vspace*{-1cm}
\caption{\label{fig_cor}
Scalar $(\bar qq)$ and $(q^TC\gamma_5q)$ correlation functions
for different chemical potentials.}
\vspace*{0.4cm}
\end{minipage}
\hspace{\fill}
\begin{minipage}[t]{75mm}
\epsfxsize=7cm
\epsffile{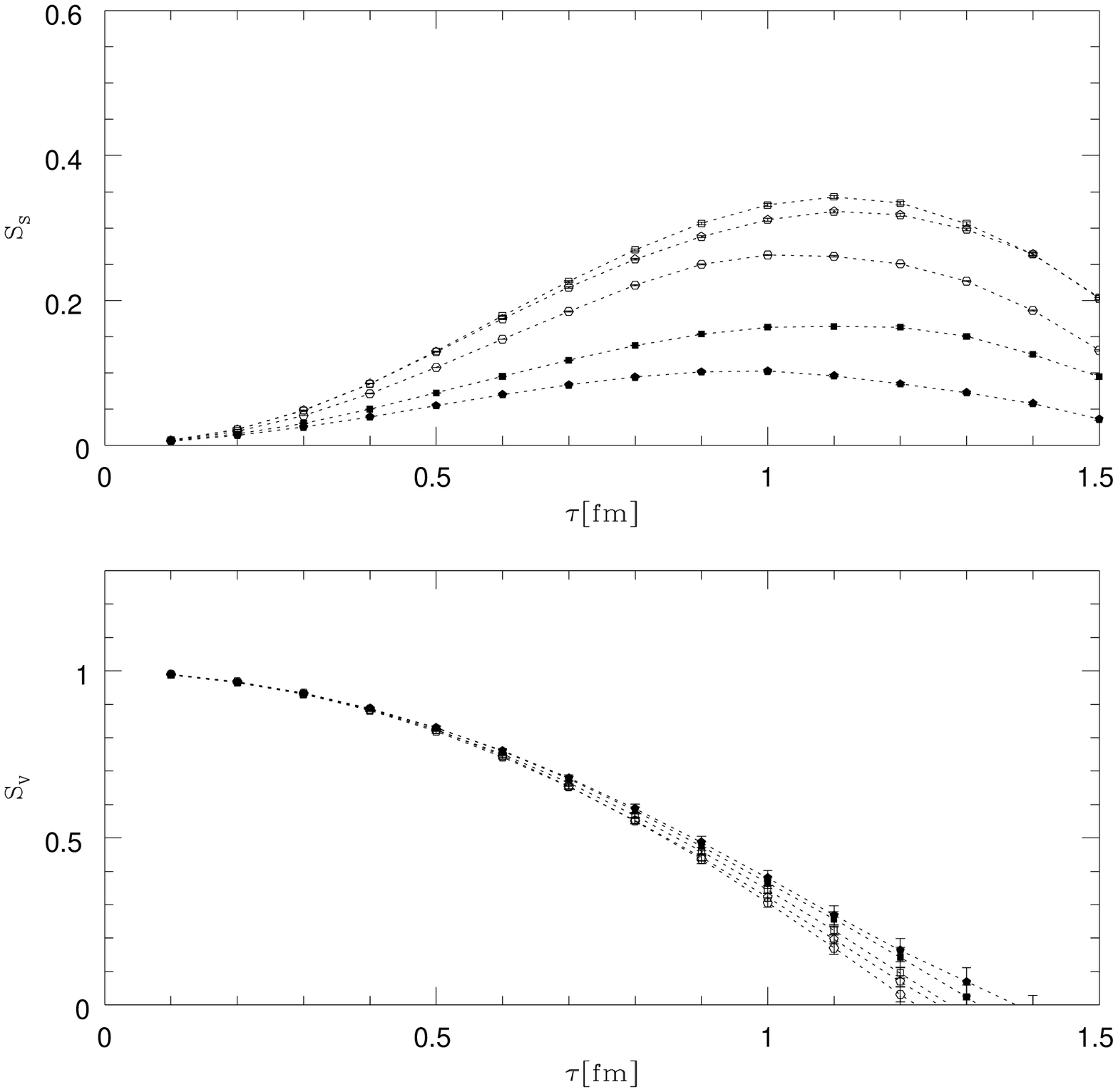}
\vspace*{-1cm}
\caption{\label{fig_prop}
Scalar $Tr(S)$ and vector $Tr(\gamma_0S)$ components of the quark 
propagator for different chemical potentials.}
\end{minipage}
\end{figure}

\section{QCD with two colors}

   A nice model system in which we can study diquark condensation 
is QCD with 
two colors. $N_c=2$ QCD has a particle-anti-particle (Pauli-G\"ursey) 
symmetry. This symmetry implies that mesons and diquarks are 
degenerate, and that (in the limit of zero mass) the quark-anti-quark
condensed state is equivalent to a diquark condensed state. If 
the quark mass is finite, the degeneracy is lifted and the true
ground-state has a chiral condensate. If we now also turn on a 
chemical potential, we expect the order parameter to rotate in
the diquark direction as soon as $\mu$ exceeds a critical value
on the order of scalar diquark mass. 

   This phenomenon can be studied in imaginary time simulations, 
because the fermion determinant remains real even in the presence
of a chemical potential. As an example, we consider the instanton
liquid simulations described in \cite{Sch_97}. The chiral condensate
is easy to measure, and we observe the expected drop at large $\mu$. 
The diquark condensate cannot be studied directly, but we can 
measure the diquark correlation function. If diquarks are condensed,
the correlator will tend to a finite value at large distance. This
is indeed observed in the results shown in Fig. 4. (In the case of
the diquark correlator, the lower points correspond to small $\mu$,
the upper points to large $\mu$. For the $(\bar qq$) correlator, the
situation is reversed.)
 
  We can also study the quark propagator in more detail. The scalar
(chirality violating) component disappears as the chemical potential
is increased, see Fig. 5. But the vector part changes very little,
indicating that the gap in the spectrum remains, even though the
quark condensate goes to zero. 

\section{Summary}

  Straightforward arguments suggest that cold quark matter is a
superconductor. Instanton effects can lead to sizeable gaps, and
we estimate the maximum gap to be on the order of 50 MeV. The 
phase structure of the condensed state, in particular if strange 
quarks are included, is very rich, and many interesting phenomena 
remain to be explored.

\end{document}